\documentclass[aps,11pt,prd,superscriptaddress,preprintnumbers, 
	amsfont,
	amssymb,
	amsmath,
	notitlepage,
	longbibliography,
	nofootinbib]{revtex4-1}
\usepackage[colorlinks=true, 
	linkcolor=blue, 
	citecolor=blue, 
	urlcolor=blue, 
	linktocpage=true]{hyperref}
\usepackage[english]{babel}
\usepackage{graphicx}

\begin{document}

\title{Notes on false vacuum decay in quantum Ising models}
\date{\today} 

\author{Ian G. Moss}
\email{ian.moss@newcastle.ac.uk}
\affiliation{School of Mathematics, Statistics and Physics, Newcastle University, 
Newcastle Upon Tyne, NE1 7RU, UK}

\begin{abstract}
This paper aims to gather together some of the basic ideas behind the theory of
false vacuum decay in quantum Ising models, focusing on the application
of spin chains as analogue systems to false vacuum decay in elementary particle
theory. Elementary results on quantum Ising models are reformulated and extended
to more closely resemble the theory of false vacuum decay in quantum field theory.
The dynamics of bubble walls is investigated, and
a speculative conjecture for the false  vacuum decay rate in a two 
dimensional quantum Ising model
is put forward.
\end{abstract}

\maketitle

\section{Introduction}

False vacuum decay \cite{Coleman:1977py,Callan:1977pt} is a fascinating 
prediction of elementary particle physics. Broadly
speaking, false vacuum decay is the decay of a large-scale metastable state
in a quantum field theory. The decay takes place by the nucleation of regions
of a stable phase, known variously as droplets, bubbles, bounces or instantons.
From a theoretical perspective, the process is a type of non-equilibrium, 
non-perturbative quantum phenomena. It is truly remarkable that the process
has a fairly simple theoretical description in terms of certain critical field
configurations. 

Real-world realisations of false vacuum decay would only happen in the most 
extreme conditions in the early universe \cite{Linde:1981zj,Hawking:1982ga} 
or in the speculative prospect of Higgs decay 
\cite{Coleman:1980aw,Turner:1982xj,Burda:2015isa}.  
Analogue systems have opened up the possibility of 
exploring false vacuum decay in a whole new range of physical systems, 
from cold atoms 
\cite{FialkoFate2015,Braden:2017add,Billam:2018pvp,Braden:2018tky,Billam:2021nbc,Jenkins:2023npg,Billam:2023} 
to quantum spin chains 
\cite{PhysRevB.104.L201106,PhysRevLett.133.240402}, and 
experimental studies of false vacuum decay are now being realised
\cite{Zenesini:2024,Vodeb_2025}.

The range of analogue systems is broad. On the one hand, the system should help
verify (or falsify) some of the conclusions that have been reached about the original system, for
example how false vacuum decay in the early universe could produce topological defects.
Beyond this, the analogue may offer new insights that are not covered by existing theory,
for example how the positions of structures produced by bubble collisions might be correlated. 
The final aspect of analogue systems is the existence of interesting physical processes that
go beyond the analogue. These are often the most difficult to analyse, as they
are likely to involve non-linear, many-body quantum phenomena.

This paper aims to gather together some of the basic ideas behind the theory of
false vacuum decay in quantum Ising models, focusing on the analogue systems aspect. 
Along the way, we will try to rewrite
quantum spin chain results in a way that more closely resembles the literature on false 
vacuum decay, and bring out features that are most relevant for an analogue
to field theory. 
The new results in this paper are on the dynamics of bubble walls, new field
theory formulations and a speculative conjecture 
for the false vacuum decay rate in a two dimensional quantum Ising model.

The quantum Ising model consists of spins arranged on the nodes of a fixed spatial
lattice, with nearest neighbour interactions between the spins. Ferromagnetic 
states are where the spins are all aligned in one direction, specifically the up or down direction. 
One of these ferromagnetic states can be metastable 
when there are forces both in the direction perpendicular and parallel to the
direction of magnetisation. Although the theoretical description
is not a quantum field theory,  the nodes can be thought 
of a lattice approximation to the spatial continuum, and the spins act in place of the field space.

The quantum Ising model can be reformulated in various ways. (For a review, see e.g. \cite{Suzuki2013}.)
In the first place, for the quantum Ising model in one dimension, there is a precise reformulation in terms of 
fermions on a lattice \cite{RevModPhys.51.659}. On a less rigorous footing, there is a link between the 
quantum model and a classical Ising model in one higher spatial dimension \cite{PhysRevD.17.2637}. 
 In a pioneering paper, Rutkevich \cite{Rutkevich_1999} derived the false vacuum decay rate and proved
consistency of the results using both approaches. The decay rate
in one dimension for a system with $N$ spins is
\begin{equation}
\Gamma_{\rm nuc}=\frac{\pi}{9}h\, Ne^{-q/h},\label{Rut}
\end{equation}
where $h$ is the (half) difference in energy between the metastable
and true vacuum, and $q$ is a known function of the magnetisation
and transverse coupling of the spin chain. The numerical factor in front of the exponential
has so far only been obtained from the fermion representation.
Note that the units have been chosen such that $\hbar=1$.

Recent theoretical work on the one dimensional system has explored
the quantum theory on finite chains by means of numerical simulations
\cite{PhysRevB.104.L201106,PhysRevLett.133.240402,Maki2023}.
The numerical results in \cite{PhysRevB.104.L201106} are consistent
with the theoretical prediction (\ref{Rut}), with $h$ of order $10^{-2}$
and transverse coupling around $0.7$ times the magnetisation.
Refs. \cite{PhysRevLett.133.240402} and \cite{Maki2023} have
explored the fermion representation of the quantum Ising model
in greater detail than  \cite{Rutkevich_1999}, examining
questions about the pre-factor in the nucleation rate.
Numerical simulations of two dimensional systems have also been performed 
\cite{borla2026,Pavesic:2026yiz}, as have some preliminary experimental 
explorations \cite{osterholz2026,humar2026}. 
The lack of a fermion representation means a rigorous 
prediction for the false vacuum decay rate in this case is lacking. 
We make a conjecture for a possible rate in the case
of small transverse coupling  in a section V.

The plan of this paper is as follows. Section II develops the theory of false vacuum decay.
Section III describes the motion of the bubble wall in a quantum spin system.
Field theory reformulations of the spin system are introduced in section IV, and
higher dimensional spin systems are discussed in section V.

\section{False vacuum decay on the spin chain in one dimension}

We start with the quantum Hamiltonian for a system that undergoes false vacuum decay,
\begin{equation}
H=-J\sum_{j=1}^{N}S_j^zS_{j+1}^z-gJ\sum_{j=1}^{N}S_j^x-hJ\sum_{j=1}^{N}S_j^z.
\end{equation}
The ${\bf S}_j$ are Pauli spin operators normalised to have eigenvalues $\pm1$, located along a circular spin chain.
The coefficient $J$ is the magnetic coupling strength between sites. The longitudinal coefficient $h$ gives half the difference
in energy between the $\pm 1$ states, and the transverse coupling coefficient $g$ allows 
for transitions between different vacua.

The eigenvectors of $S_j^z$ with eigenvalues $S_j$ form a convenient basis of states,  $|S_j\rangle=|S_1,S_2,\dots S_{N}\rangle$.
The fully magnetised states $|TV\rangle=|1,1,\dots 1\rangle$ and 
$|FV\rangle=|-1,-1,\dots \rangle$ are identified as the true and false vacua
in the limit that $g\to 0$ (and $\epsilon>0$). The states  $|S_j\rangle$ are no longer stationary states when $g\ne0$,
so that transitions from the false vacuum take place. The large density of states for the spin system
is important in ensuring that the system will almost always be close to the true vacuum at late times
if the energy difference $h$ is small. We call this false vacuum decay. This is in contrast to a purely quantum 
mechanical tunnelling problem with a 
small number of degrees of freedom where there are recurrences back to the false vacuum.

\subsection{Decay rates}

Following Coleman's approach to false vacuum decay  \cite{Coleman:1977py,Callan:1977pt}, 
we identify the lowest energy 
eigenvalue $E_0$ using a large imaginary time limit of the vacuum amplitude,
\begin{equation}
\langle FV | e^{-HT/\hbar}| FV\rangle\to Ae^{-2E_0 T/\hbar}\hbox{ as }T\to\infty\label{limit}.
\end{equation}
The decay rate $\Gamma_{\rm nuc}$ is associated with having an imaginary part to the energy eigenvalue,
\begin{equation}
\Gamma_{\rm nuc}=\frac{|{\rm Im}E_0|}{\hbar}\label{rate}.
\end{equation}
We calculate the amplitude by introducing a path integral along the lines of the approach described in Ref. \cite{Suzuki2013}.

In order to turn the matrix element into a path integral, we split the time interval into segments of time $\delta\tau$,
\begin{equation}
\langle FV | e^{-HT/\hbar}| FV\rangle=\sum_{S_{jk}}\langle FV|e^{-H\delta\tau/\hbar}|S_{j1}\rangle\langle S_{j1}|\dots
|S_{jN}\rangle\langle S_{jN}|e^{-H\delta\tau/\hbar}|FV\rangle.
\end{equation}
Consider the contribution from a single segment. In general, we cannot split the exponential term into two parts, but for
small $\delta\tau$ we can approximate by
\begin{equation}
\langle S_{jk}|e^{- H\delta\tau /\hbar}|S_{jk+1}\rangle
=\sum_{S'_{j}}\langle S_{jk}|e^{J_1\sum S_i^zS_{i+1}^z+\frac12\lambda\sum S_j^z}|S'_{j}\rangle
\langle S'_{j}|e^{\Gamma\sum S_i^x}|S_{jk+1}\rangle,
\end{equation}
where we introduce parameters
\begin{equation}
J_1=J\delta\tau,\quad\lambda=2hJ\delta\tau,\quad\Gamma=gJ\delta\tau. 
\end{equation}
The first factor is simple to evaluate,
\begin{equation}
\langle S_{jk}|e^{J_1\sum S_j^zS_{j+1}^z+\frac12\lambda\sum S_j^z}|S'_{j}\rangle
={\rm exp}\left\{J_1\sum S_{jk}S_{j+1k}+\frac12\lambda \sum S_{jk}\right\}\delta_{SS'}.
\end{equation}
For the second factor, the matrix elements of $S^x$ for a single spin are
\begin{equation}
\langle S'|(S^x)^n|S\rangle=\frac12(1+(-1)^nSS'),\label{sx}
\end{equation}
where $S,S'=\pm1$.
Taking the exponential, noting that the right hand side of (\ref{sx}) are projection matrices, gives
\begin{equation}
\langle S_{jk}|e^{\Gamma\sum S_i^x}|S_{jk+1}\rangle=
2\,{\rm exp}\left\{
\frac12\sum(1+S_{jk}S_{jk+1})\ln\cosh\Gamma+\frac12\sum(1-S_{jk}S_{jk+1})\ln\sinh\Gamma
\right\}.
\end{equation}
Combining the two amplitudes, and dropping an overall constant for the moment,
\begin{equation}
\langle S_{jk}|e^{- H\delta\tau /\hbar}|S_{jk+1}\rangle=
{\rm \exp}\left\{
J_1\sum S_{jk}S_{j+1k}+J_2\sum S_{jk}S_{jk+1}+\frac12\lambda \sum S_{jk}
\right\},
\end{equation}
where
\begin{equation}
J_2=\frac12\ln\coth\Gamma.
\end{equation}
Already, there is something troubling here. The small $\delta\tau$ limit was used earlier to split
the exponential factor, but the limit  $\delta\tau\to 0$ in $J_2$ diverges. 
We shall return to this point below.

Putting together the time segments give the full matrix element which we can write as
\begin{equation}
\langle FV |e^{-HT/\hbar}| FV\rangle=\int DS e^{-S_E/\hbar},\label{pi}
\end{equation}
where we call $S_E$ the Euclidean action,
\begin{equation}
\frac{S_E}{\hbar}=-J_1\sum_{jk} S_{jk}S_{j+1k}-J_2\sum_{jk} S_{jk}S_{jk+1}-\frac12\lambda \sum_{jk} S_{jk}.
\end{equation}
The $j$ sum runs from one to $N_1=N$, and the $k$ sum runs from one to $N_2=T/\delta\tau$. 
The measure, including the factor which we dropped earlier,  is
\begin{equation}
\int DS=\left(\frac{\sinh2\Gamma}{2}\right)^{N_1N_2/2}\sum_{S_{jk}}.
\end{equation}
A useful way of the rewriting the action $S_E$ to look more like a field theory is
\begin{equation}
\frac{S_E}{\hbar}=\sum_{j}\left\{
\frac12J_1(D_x S)_j^2+\frac12J_2(D_y S)_j^2
-\frac12\lambda S_{j}
\right\},\label{action}
\end{equation}
where $j$ runs through a two-dimensional lattice and $D_x$, $D_y$ are forward difference operators
in the respective directions, $(D_yS)_{jk}=S_{jk+1}-S_{jk}$. There is a constant shift in the action when written in this form
which changes the measure to become
\begin{equation}
\int DS=\left(\sinh\Gamma\right)^{N_1N_2/2}\sum_{S_{jk}}.
\end{equation}

We need to return to the limit $\delta\tau\to0$. The divergence of $J_2$ is not a flaw
in our derivation, as other methods, such as the ``transfer matrix'' methods used in 
\cite{PhysRevD.17.2637,RevModPhys.51.659}, lead to the same conclusion.
The physical explanation is that the correlations functions of the spin chain for fixed time
must remain fixed as $\delta\tau\to0$, but the number of spins in the time direction diverges.
The two dimensional Ising model achieves this by becoming strongly asymmetric in this limit.
In terms of false vacuum decay, the consequence is that we have to take the $T\to\infty$
limit before taking the $\delta\tau\to0$ limit.

\subsection{Vacuum droplets I}

The next task is to evaluate the imaginary part of the lowest energy eigenvalue using the
path integral formula (\ref{pi}). The simplest approach is to look for an extremum
of the path integral. Though we shall see that this is extremely naive, it is nevertheless quite
instructive.

\begin{figure}[htb]
\begin{center}
\vskip 0.5cm
\scalebox{0.1}{\includegraphics{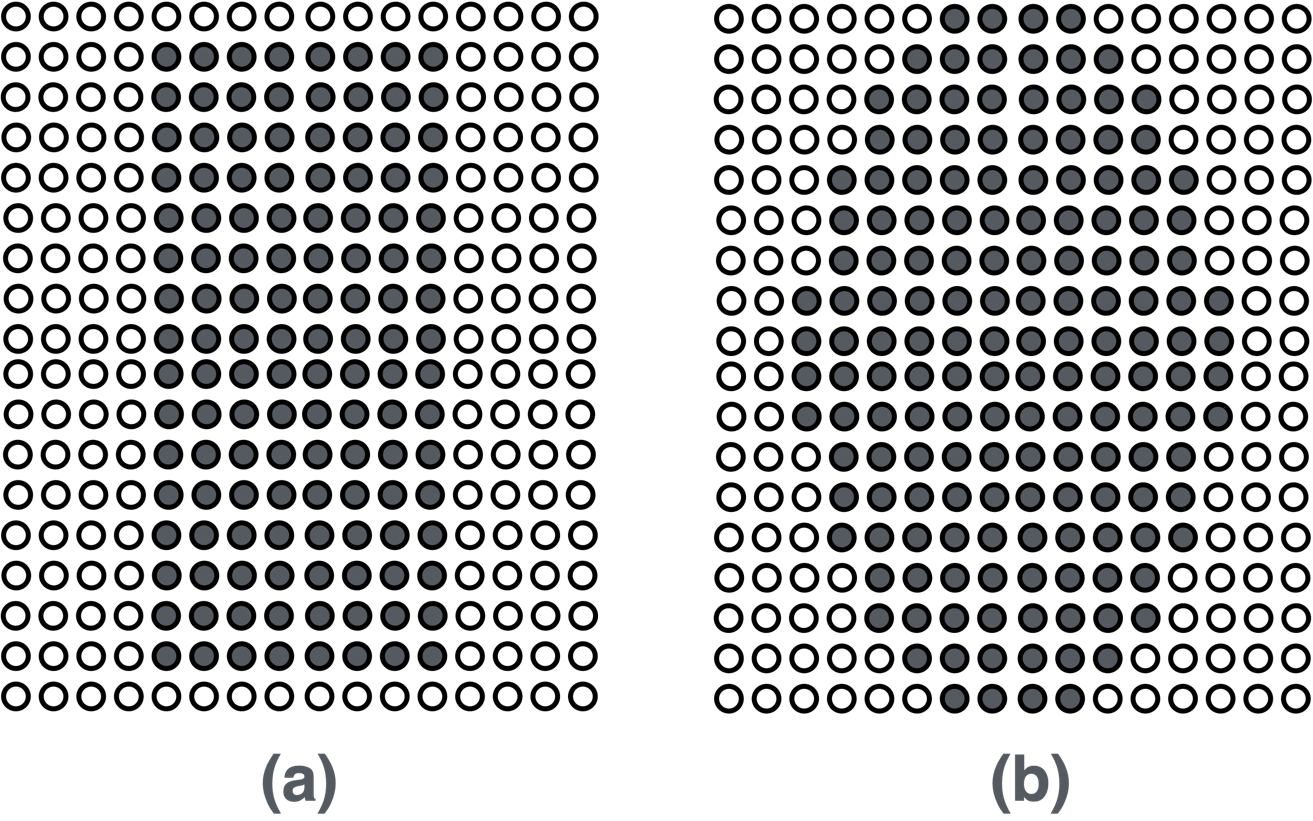}}
\end{center}
\caption{
(a) Rectangular and (b) ellipsoidal droplets with filled circles representing spin up. 
The ellipsoidal contribution to the action for
lengths of the axes that make the action stationary is larger than the rectangular 
contribution by $4/\pi$.
}
\label{fig:droplet}
\end{figure}

\begin{figure}[htb]
\begin{center}
\vskip 0.5cm
\scalebox{0.1}{\includegraphics{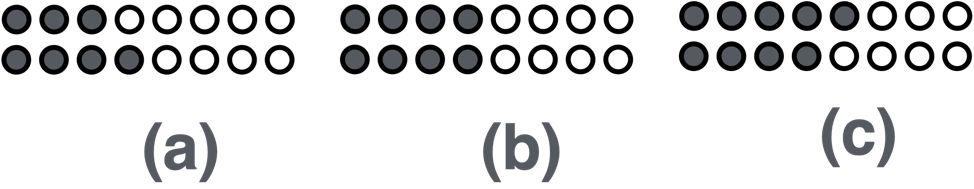}}
\end{center}
\caption{
Two adjacent rows of spins.
The three possibilities, where the edge of the droplet has $dy=1$ and (a) $dx=-1$, 
(b) $dx=0$ and (c) $dx=1$, all contribute $2J_1|dy|+2J_2|dx|$ to the action.
}
\label{fig:spins}
\end{figure}

To start with, consider a droplet region $D$ of true vacuum spins with spins $S_{jk}=1$, 
surrounded by false vacuum $S_{jk}=-1$, as in figure \ref{fig:droplet}. 
Define the shifted action $S_E(D)$ by subtracting
off the action of the false vacuum,
then from (\ref{action}),
\begin{equation}
S_E(D)=\int_{\partial D}(2J_1|dy|+2J_2|dx|)-\lambda\int_Ddxdy,\label{dropaction}
\end{equation}
where $x$ and $y$ lie on a grid with unit spacing (see Fig. \ref{fig:spins}). 
For a rectangular region $\square$ with edges of size $n_1$ and $n_2$,
\begin{equation}
S_E(\square)=4J_1n_2+4J_2n_1-\lambda n_1n_2.\label{B}
\end{equation}
If the edges of the rectangle are distorted, whilst keeping the area fixed, then the action
increases. The rectangle is a minimum of the action under these distortions. If we change 
$n_1$ or $n_2$, we obtain a stationary point for
\begin{equation}
n_1=\frac{4J_1}{\lambda},\qquad n_2=\frac{4J_2}{\lambda},
\end{equation}
with action
\begin{equation}
S_E(\square)=\frac{16J_1J_2}{\lambda}.
\end{equation}
The actual edge lengths should be integers, but the difference between the real and integer values
is small when $\lambda\ll J_1$ and the rectangles contain a large number of spins.
The second derivative at the stationary point has eigenvalues $\pm\lambda$, 
revealing that the rectangle is a minimum 
of the action under skew distortions and a maximum of the action under dilations. 
This saddle point situation is generic for a critical droplet theory, and in the context of false vacuum 
decay it is the negative mode of perturbations about the critical droplet that generate an imaginary 
part to the energy eigenvalue. We identify the rectangle at this stationary point as the critical droplet.

The full path integral should take into account the fluctuations about the critical
droplet as well as including contributions from multiple droplets. Suppose
a single droplet contributes a factor $(i\lambda)^{-1} Ke^{-S_E/\hbar}$, with 
$i\lambda=\sqrt{(-\lambda)(\lambda)}$ from integrating out the skew
distortions and the dilations, and the remaining perturbations contributing a factor $K$.
Placing ``$n$'' identical droplets in an area $N_1N_2$, but ignoring their overlaps, and summing,
gives a total contribution of
\begin{equation}
\sum_{n=0}^\infty\frac{1}{n!}(N_1N_2)^n(-i\lambda^{-1} Ke^{-B})^n=\exp(-iN_1N_2\lambda^{-1} Ke^{-S_E/\hbar}).
\end{equation}
From (\ref{limit}) and (\ref{rate}), we read off the decay rate
\begin{equation}
\Gamma_{\rm nuc}=\frac{N_1N_2 K}{2 \lambda T}\,e^{-B}=\frac{N}{4hJ\delta\tau^2}Ke^{-S_E/\hbar},
\end{equation}
where we have used $N_2=T/\delta\tau$ and $\lambda=2h\delta\tau$. The action for the critical droplet is
\begin{equation}
S_E=\frac{4}{h}\ln\frac{1}{gJ\delta\tau}.
\end{equation}
It is clear that we have severe problems taking the limit $\delta\tau\to 0$.
The difficulty has arisen from an over-reliance on the classical Ising model in two dimensions.
Nevertheless, for the present, we make an arbitrary decomposition,
\begin{equation}
S_E=\frac{4}{h}\ln\frac{b}{g}
-\frac{4}{h}\ln bJ\delta\tau.\label{decomp}
\end{equation}
For the $K$ contribution we tentatively use a result obtained from a dual field theory model \cite{Langer1967,Gunther1980},
suggesting that droplet excitations give $K\propto\lambda^2$.
The decay rate resulting from dropping the final term in (\ref{decomp}) would be
\begin{equation}
\Gamma_{\rm nuc}=BNh\exp\left\{
-\frac{4}{h}\ln\frac{b}{g},
\right\}\label{nuc}
\end{equation}
where $B$ is a constant.
Remarkably, this turns out to be a limiting $g\ll 1$ form of the result given in the introduction (\ref{Rut}), as we shall see
in the next section. The naive argument has given the correct coefficient of the logarithm term in the exponent, and
the factor $1/g$ appearing inside the logarithm. The $b$ factor
inside the logarithm is the only term not accounted for in this approach.

\subsection{Vacuum droplets II}

Maybe the problems we found for critical droplet could be resolved if we
used an effective action $\Gamma_E$ that included quantum corrections
to the classical action $S_E$? Ordinarily, we would look to one loop perturbation
theory. However, this proves problematic due to the discreteness of the
spin variables. Remarkably, exact results exist for the
correlation functions of the two-dimensional Ising model for all values
of the nearest neighbour couplings. A duality relation exists between the 
correlation length $\xi$ and the surface tension
of a droplet $\sigma$ \cite{zia1978duality}. From the surface tension it is possible 
obtain an effective action for the wall of a droplet.

The detailed calculation of $\sigma$ is complicated, so we will simply
quote the expresion for the effective action $\Gamma_E[D]$ for a droplet $D$ of true vacuum 
\cite{PhysRevB.25.2042},
\begin{equation}
\Gamma_E[D]=\int_{\partial D}\left(\alpha_xdy-\alpha_ydx\right)-\lambda\int_D dxdy,
\label{Zaction}
\end{equation}
where $\alpha_x$ and $\alpha_y$ are functionals of the boundary curve $\partial D$.
These are defined implicitly by,
\begin{align}
a_x\cosh\alpha_x+a_y\cosh\alpha_y=1,\label{a1}\\
a_x dx\,\sinh\alpha_x+a_ydy\,\sinh\alpha_y=0,\label{a2}
\end{align}
where
\begin{equation}
a_x=\frac{\tanh 2J_2}{\cosh{2J_1}},\quad a_y=\frac{\tanh 2J_1}{\cosh{2J_2}}.
\end{equation}
We can find the critical droplet as follows. Parameterise the boundary by
$x(s)$ and  $y(s)$. From (\ref{a1}) and (\ref{a2}), it follows that under variation of the
boundary, $\delta\alpha_x y'-\delta\alpha_y x'=0$. Hence,
\begin{equation}
\delta\Gamma_E=\int\left(\alpha_x\delta y'-\alpha_y\delta x'-\lambda x\delta y'+\lambda y\delta x'\right)ds.
\end{equation}
Integrating by parts, we deduce that $\delta\Gamma_E=0$ when $\alpha_x'=\lambda x'$ and $\alpha_y'=\lambda y'$. 
Placing the droplet centre at the origin
gives $\alpha_x=\lambda x$ and $\alpha_y=\lambda y$. The critical droplet using (\ref{a1}) is then
\begin{equation}
a_x\cosh\lambda x+a_y\cosh\lambda y=1.
\end{equation}

\begin{figure}[htb]
\begin{center}
\scalebox{0.3}{\includegraphics{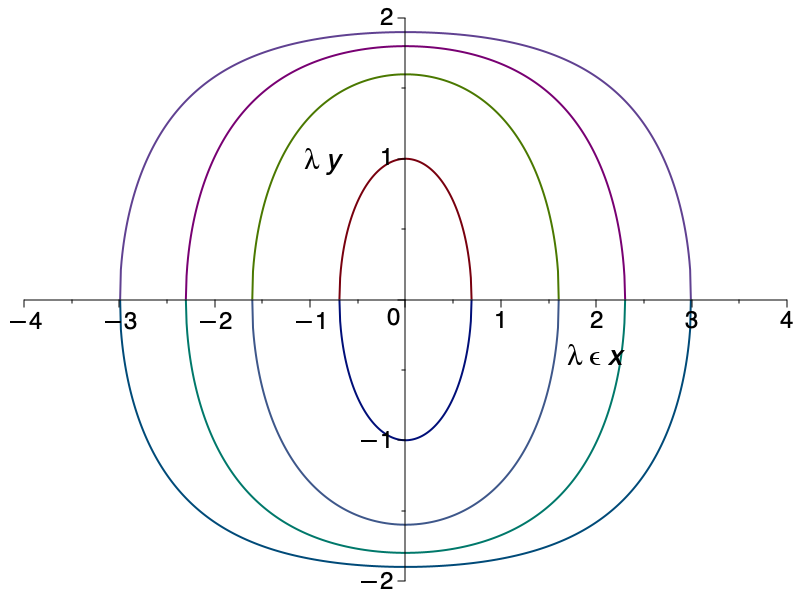}}
\end{center}
\caption{
The critical droplet for a one dimensional quantum spin chain is shown for various values 
of $J/\Gamma=2,5,10,20$ moving outwards form the centre, where $J$ is the magnetisation 
and $\Gamma$ the transverse coupling. 
The shape becomes more ``rectangular'' as $J/\Gamma$ increases.
}
\label{fig:sample}
\end{figure}

\begin{figure}[htb]
\begin{center}
\scalebox{0.3}{\includegraphics{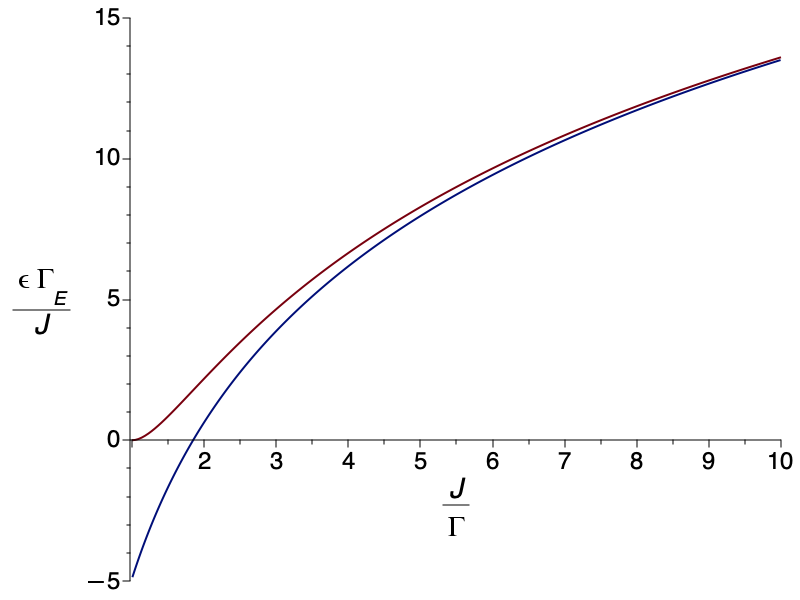}}
\end{center}
\caption{
The nucleation exponent $\Gamma_E$ for the one dimensional quantum spin chain (upper curve)
and the logarithmic approximation (lower curve), plotted against $J/\Gamma$, where $J$ is the magnetisation
and $\Gamma$ the transverse coupling.
}
\label{fig:sample}
\end{figure}

In the vacuum decay model, $J_1\ll 1$, and $J_2=(\ln\coth\Gamma)/2\gg 1$. The coefficients $a_x$ and $a_y$ become
\begin{equation}
a_x\approx1-2\Gamma^2-2J_1^2,\quad
a_y\approx4\Gamma J_1.\label{axay}
\end{equation}
For $\lambda x$ small,
\begin{equation}
a_x(1+\lambda^2x^2/2)+a_y\cosh\lambda y\approx 1,
\end{equation}
which reduces, using (\ref{axay}) and $J_1=J\delta\tau$, $\lambda=2h\delta\tau$ and $\Gamma=Jg\delta\tau$,  to
\begin{equation}
x\approx\pm\frac{1}{h}\left(1+g^2-2g\cosh\lambda y\right)^{1/2}.
\end{equation}
The $y$ range for a closed curve becomes $-\ln(1/g)<\lambda y<\ln(1/g)$. If we substitute
into the action (\ref{Zaction}), we get
\begin{equation}
\Gamma_E[D]=\frac{4}{h}\int_0^{\ln(1/g)}d(\lambda y)
\left(1+g^2-2g\cosh\lambda y\right)^{1/2}.
\end{equation}
The integral can be expressed in closed form in terms of complete elliptic integrals of the first
and second kind, $K$ and $E$,
\begin{equation}
\Gamma_E[D]=\frac{16(1+g)}{h}
\left[
K\left(\frac{1-g}{1+g}\right)-E\left(\frac{1-g}{1+g}\right)
\right].
\end{equation}
This result is independent of $\delta\tau$ and hence finite in the limit $\delta\tau\to0$.

As an aside, we note that the fermionic version of the quantum spin chain has dispersion relation
\begin{equation}
\omega(k)=\left[\left(1-g\right)^2+4g\sin^2 k\right]^{1/2}.
\end{equation}
The integrand in the exponent $\Gamma_E[D]$ is given in terms of the dispersion relation by 
$\omega(i\lambda y/2)$.
This is what was found in ref. \cite{Rutkevich_1999}. The fermions are related to spin
``kink" configurations that interpolate between up spins and down spins. Bubbles have two
kinks, and can be interpreted as fermion pairs, with an interaction depending on the 
longitudinal coupling $h$.

In the case when $g\ll 1$, the nucleation rate result reduces to
\begin{equation}
\Gamma_E\approx\frac{4}{h}\ln\frac{4}{e^2g},
\end{equation}
allowing us validate dropping the divergent $\ln(bJ\delta\tau)$ term from the action in the previous section.
Including the same pre-factors as before leads to the result for the nucleation rate (\ref{nuc}).

\section{Bubble dynamics}

The discussion of false vacuum decay in the previous section identified the decay taking place though the nucleation
of a critical droplet. Although the droplet has been defined using an imaginary time coordinate, we can obtain a picture
of droplet nucleation by analytic continuation of the time variable \cite{Coleman:1977py,Callan:1977pt}.

We $y$ coordinate has spacing $\delta \tau$ in the imaginary time direction. We introduce real time $t$
though continuation to $y=i t/\delta\tau$. When we use $\lambda=2hJ\delta\tau$, the droplet
solution becomes
\begin{equation}
x=\frac{1}{h}\left((1-g)^2+4g\sin^2ht\right)^{1/2}.
\end{equation}
The droplet nucleates at time $t=0$ with  $n_c=x(0)$ spins, where
\begin{equation}
n_c=\frac{(1-g)}{h},
\end{equation}
The droplet expands, reaching a maximum size $(1+g)/h$ before collapsing.

In relativistic field theory, a true vacuum bubble nucleates with a critical radius $(d-1)\sigma/\epsilon$
in $d$ dimensions, where $\sigma$ is the surface tension of the bubble wall. The bubble wall grows
monotonically with constant acceleration to reach relativistic velocities. 
Likewise, at least for early times, we can rewrite the spin system droplet solutions in relativistic form as
\begin{equation}
x^2-c^2t^2=n_c^2,
\end{equation}
where the wall velocity (in nodes per second) asymptotes to $c=4J\sqrt{g}$.

It is interesting to note that, as for the spin system, 
bubbles in cold atom analogue systems also reach a maximum radius and then start to collapse \cite{Billam:2018pvp}.
This is known to be associated with the breakdown of Lorentz invariance of the cold atom
analogues. The analogues are only valid above a certain length scale, but the
Lorentz contraction of the bubble wall takes the system beyond the acceptable range. Analogue systems 
with a wider range of validity expand for longer before collapse.

So far, the bubble under consideration has been the one responsible for nucleation events.
There are other situations, for example when a bubble seed is placed in the initial configuration,
where it would be desirable to have an analytic form for the evolving bubble wall. The necessary
theory can be found by constructing a Hamiltonian formulation from the action given in the 
last section, Eq. (\ref{Zaction}). If the walls are at $\pm x$, the Hamiltonian is simply 
$H=\omega(p)-hx$, where $\omega(p)$ is the fermion dispersion relation given previously. 
The Hamiltonian is constant for bubble wall trajectories,
implying they have the same shape as the critical bubble, but with left and right hand walls 
shifted by a constant amount. This is a special feature of the 1D quantum Ising model,
in which the wall motion is independent of the size of the interior. Moving beyond one dimension
gives a different scenario in which the motion of the wall depends very much on the bubble size.
We will touch on this feature in the last section.

\section{Scalar field theory}

We have quoted results for the pre-factor that were obtained from scalar field theory.
We shall take a look now at the reasoning behind this, loosely adapting the work by Langer in 
an appendix of Ref. \cite{Langer1967}. The approach has serious problems,
but is worth reviewing because it gives the correct results for some factors in the 
nucleation rate. We will argue in the next section that a complex field theory is 
a closer analogue to the quantum Ising model.

To begin with, we use the notation $Z[H]$ for the vacuum amplitude, allowing for
an external source $H_j$. The results of the earlier section imply
\begin{equation}
Z[H]=\int dS e^{-S_E+\sum H_jS_j},
\end{equation}
where
\begin{equation}
S_E=\sum_j
\left\{
\frac12J_1(D_xS)_j^2+\frac12J_2(D_yS)_j^2-\frac12\lambda S_j
-\frac12\alpha S_j^2
\right\}.
\end{equation}
The last term with coefficient $\alpha$ is a new addition. It is a constant term because $S_j^2=1$,
and we can absorb this into the measure.
Note that the energy eigenvalue we introduced earlier
is related to $\log Z[0]$. In a thermal system, this would be the
free energy, and so the decay rate would be given by the imaginary part
of the free energy. However, our
primary interest is in the decay rate at zero temperature.

We rewrite the action in matrix form as
\begin{equation}
S_E=\frac12\sum_{jk}J_{jk}S_jS_k-\frac12\sum_{j}\lambda S_j,
\end{equation}
where the matrix ${\bf J}=J_{jk}$ can be written in terms of second order difference
operators $D_x^2$ and $D_y^2$,
\begin{equation}
{\bf J}=-\alpha{\bf I}-J_1D_x^2-J_2D_y^2.
\end{equation}
We now introduce an integral transform with a real-valued field $\phi_j$ defined on the spin lattice,
\begin{equation}
Z[0]=(2\pi)^{N/2}({\rm det}|{\bf J}|)^{-1/2}\left(\prod_i\int d\phi_i\right)
\int dS\exp\left\{
\frac{1}{2}\sum_{ij}(J^{-1})_{ij}\phi_i\phi_j+\sum_i(\lambda/2+\phi_i)S_i
\right\}.
\end{equation}
Summing over the spins
\begin{equation}
Z[0]=\int D\phi
\exp\left\{
\frac{1}{2}\sum_{ij}(J^{-1})_{ij}\phi_i\phi_j+\sum_i\ln\cosh(\lambda/2+\phi_i)
\right\},
\end{equation}
where
\begin{equation}
\int D\phi=\left(8\pi e^{\alpha}\sinh\Gamma\right)^{N/2}({\rm det}|{\bf J}|)^{-1/2}\prod_id\phi_i.
\end{equation}
Convergence of the integral requires that the matrix ${\bf J}$ is negative definite. Considering
field configurations with $\phi_j\propto(-1)^{j_xj_y}$ shows that this
imposes a condition $\alpha>2(J_1+J_2)$.

If we work with a gradient expansion, then the leading terms in the inverse of the matrix ${\rm J}$
are
\begin{equation}
({\bf J})^{-1}=-\frac{1}{\alpha}{\bf I}+\frac{J_1}{\alpha^2}D_x^2+\frac{J_2}{\alpha^2}D_y^2.
\end{equation}
Inserting these into the scalar field model (and restoring the source $H$) gives
\begin{equation}
Z[H]=\int D\phi e^{-S_E[\phi]+\sum H\tanh(\phi)},
\end{equation}
where
\begin{equation}
S_E[\phi]=\sum_j\left\{
\frac{J_1}{2\alpha^2}(D_x\phi)^2+\frac{J_2}{2\alpha^2}(D_y\phi)^2+V(\phi)
\right\},
\end{equation}
and the potential $V$ is
\begin{equation}
V(\phi)=\frac{\phi^2}{2\alpha}-\ln\cosh(\phi+\lambda/2).
\end{equation}

The potential has metastable states only for $\alpha>1$. The instanton
approach can be used to obtain a vacuum nucleation rate as before, but
the result depends on the undetermined parameters $\alpha$ and $\delta\tau$.
Taking the continuum limit is problematic.
Nevertheless, the dependence on the
parameter $\lambda$ is unambiguous. The contribution from ``Goldstone''
modes gives a factor $\propto \lambda^2$ in the pre-factor \cite{Langer1967,Gunther1980}. 
Together with from the zero modes (implicitly included already in the earlier discussion), gives
\begin{equation}
\Gamma_{\rm nuc}=B\lambda\exp(-A/\lambda),
\end{equation}
for $A$ and $B$ independent of $\lambda$, which is consistent with the earlier results, 
and used by Rutkevich in \cite{Rutkevich_1999}.

Part of the problem with the field theory approach can be seen if we take the kinetic
term in a continuum
field theory and discretise the imaginary time in units of $\delta\tau$,
\begin{equation}
\int\frac12(\partial_\tau\phi)^2d\tau\approx\frac{1}{\delta\tau}\sum_j\frac12(D_y\phi)^2.
\end{equation}
The coefficient $J_2$ should diverge as $1/\delta\tau$, but instead it is
logarithmically divergent. This problem would be far less severe if the
kinetic term was first order in derivatives, as in a non-relativistic theory.

\subsection{Non-relativistic field theory}

It is possible to introduce a non-relativistic theory with a complex boson field. The identification
to the quantum Ising model is still problematic (at least in the simple version presented below),
but unlike the fermion dual, this transformation can be done in any dimension. 

We can rewrite the Ising model action with first order derivatives,
\begin{equation}
S_E=\sum_j
\left\{
\frac12J_1(D_xS)_j^2-J_2S_j(D_yS)_j-\frac12\lambda S_j
-\frac12\alpha S_j^2
\right\},
\end{equation}
where $D_y$ is the forward difference operator in the $y$ direction. It is important here
that  $D_y^\dagger\ne -D_y$, otherwise the $D_y$ term would vanish.
We now have a non-hermitian coupling matrix ${\bf J}$,
\begin{equation}
{\bf J}=-\alpha{\bf I}-\frac12J_1D_x^2-J_2D_y
\end{equation}
In the small derivative limit, this has inverse,
\begin{equation}
{\bf J}^{-1}=-\frac{1}{\alpha}{\bf I}+\frac{J_1}{2\alpha^2}D_x^2+\frac{J_2}{\alpha^2}D_y
\end{equation}
We can use a matrix identity for a complex scalar field $\psi$,
\begin{equation}
\frac12(\bar\psi-SJ){\bf J}^{-1}(\psi-JS)=
\frac12\bar\psi{\bf J}^{-1}\psi+\frac12(\bar\psi+\psi)S+\frac12S{\bf J}S,
\end{equation}
When we insert this
identity into the path integral, the shifts in $\bar\psi$ and $\psi$ are not complex conjugates
because ${\bf J}$ is not hermitian. (However, the shift is still legal.
Rewrite the integral over $\bar\psi$ and $\psi$  in terms of integrals over the real and 
imaginary parts of $\psi$.
Move the contours over these real parts into the complex plane,
and apply Cauchy's theorem. This shifts the fields as required.)

Following the same steps as in the previous section, gives the path integral
\begin{equation}
Z[H]=\int D\bar\psi D\psi e^{-S_E[\psi]+\sum H\tanh[(\bar\psi+\psi)/2]}
\end{equation}
where
\begin{equation}
S_E[\psi]=\sum_j\left\{-\frac{J_2}{\alpha^2}\bar\psi_j(D_y\psi)_j
+\frac{J_1}{2\alpha^2}(D_x\bar\psi)_j(D_x\psi)_j+V(\psi)\right\}.
\end{equation}
Note that the symmetry between $\psi$ and $\bar\psi$ means we can replace the forward
difference operator $D_y$ by the symmetric difference operator. The potential
\begin{equation}
V(\psi)=\frac{\bar\psi\psi}{2\alpha}-\ln\cosh\frac12(\bar\psi+\psi+\lambda)
\end{equation}
The potential has a minimum for $\psi=\bar\psi\approx\pm\alpha$. For instanton solutions,
it is necessary to treat $\bar\psi$ and $\psi$ as independent fields \cite{Billam:2018pvp}.
These are harder to find than in the relativistic case due to the lack of symmetry
in the $x,y$ plane, so we leave this for anyone who might be interested.
As with the real scalar case, the limit $\delta\tau\to0$ is problematic.

\section{False vacuum decay on the spin chain in higher dimensions}

Extending the known results for the quantum spin chain in one dimension to
two dimensions is rather difficult.
We no longer have an equivalent Fermion theory to fall back on. Indeed, this seems
to be disallowed by the spin statistics theorem. We are also missing an effective
theory for the three dimensional Ising model at the present time. Although one can
apply critical droplet theory to the Ising model in arbitrary dimensions, as was done 
by Langer \cite{Langer1967} and  G\"unther \cite{Gunther1980}, we are missing an 
analytic expression for the surface tension of the droplet. 

On the other hand, we have seen that the surface tension has a simple form
in two dimensions.
In this section we make a similar conjecture to obtain the vacuum decay rate in the $d-1$
dimensional quantum spin chain.

We introduce the Hamiltonian on a lattice in $d-1$ dimensions
\begin{equation}
H=-J\sum_{<ij>}S_{i}^zS_j^z-Jg\sum_{i}S_i^x-Jh \sum_{i}S_{i}^z.
\end{equation}
The derivation of the path integral goes through almost exactly in the same way as for
the one dimensional case. The action becomes,
\begin{equation}
S_E=\sum_j\left\{
\frac12J_1(DS)_j^2+\frac12J_d(D_\tau S)_j^2
-\frac12\lambda S_j\label{3daction}
\right\},
\end{equation}
where $j$ runs over a $d$ dimensional lattice, $D$ and $D_\tau$ are 
forward difference operators in the space and imaginary time directions.
The coefficients are $J_1=J\delta\tau$ and $J_d=(\ln\coth Jg\delta\tau)/2$.

\subsection{Vacuum droplets I}

We start with a naive approach, based on the classical action in three dimensions, 
with couplings $J_1\dots J_3$.
Consider a cylinder $C$ with $n_1\approx n_2$ spins along the diameters, and 
$n_3$ spins along the length. It has shifted action $S_E(C)$,
\begin{equation}
S_E(C)=\pi J_1n_2n_3+\pi J_2n_1n_3+\pi J_3n_1n_2-\frac14\lambda \pi n_1n_2n_3.
\end{equation}
There is a saddle point at $n_i=8J_i/\lambda$, $i=1,2,3$, with action
\begin{equation}
S_E(C)=
\frac{64\pi J_1J_2J_3}{\lambda^2}.
\end{equation}
The eigenvalues of the Hessian matrix in the large $J_3$ limit
are $\pm \pi J_3$ and $2\pi J_1J_2/J_3$. The contribution to the path integral from a single droplet which
includes the Hessian matrix eigenvalues and small perturbations is now
$(-64\pi J_1J_2J_3)^{-1/2}Ke^{-S_E}$. The sum over many droplets is performed as before to give the nucleation rate,
\begin{equation}
\Gamma_{\rm nuc}=\frac{N_1N_2N_3}{8T\sqrt{\pi J_1J_2J_3}}Ke^{-S_E/\hbar}.
\end{equation}
Using $N_3=T/\delta\tau$ and the coefficients $J_1$, $J_2$, $J_3$ given above leaves
\begin{equation}
\Gamma_{\rm nuc}=\frac{N_1N_2}{8\delta\tau^2J}
\left(\ln\frac{1}{Jg\delta\tau}\right)^{-1/2}Ke^{-S_E},
\end{equation}
where
\begin{equation}
S_E=\frac{16\pi}{h^2}\ln\frac{1}{Jg\delta\tau}.
\end{equation}
We now conjecture that the same subtractions of $\ln(J\delta\tau)$ terms used in the one dimensional 
case works for the two dimensional case. For the pre-factor, we  appeal
to field theory models, but in three dimensions there are contradictory results in the literature. To allow for
this uncertainty we take $K\propto h^aJ^2$. The conjectured decay rate is
\begin{equation}
\Gamma_{\rm nuc}=B N_1N_2h^aJ\left(\ln\frac{1}{g^2}\right)^{-1/2}\exp
\left\{-
\frac{16\pi}{h^2}\ln\frac{b}{g}
\right\}.\label{3Ddecay}
\end{equation}
where $B$ is a constant (different from the one-dimensional case) and $b$ is another undetermined constant.
As an example, the results of G\"unter et al. \cite{Gunther1980} have $a=2/3$ from Goldstone mode excitations.
The one-dimensional model had $b\approx0.5$, and we would expect something similar here.
The conjecture amounts to predicting the coefficient $16\pi/h^2$, which could be tested in 
experiments which use a range of $h$.

\subsection{Vacuum. droplets II}

Next we will attempt to generalise the quantum corrections to the surface tension.
Consider a droplet $D$ of true vacuum with surface ${\bf x}(s)$ in $d$ dimensions.
We assume the surface tension takes the form ${\bf n}\cdot\boldsymbol{\alpha}$,
where ${\bf n}$ is the unit normal and $\boldsymbol{\alpha}$ is a functional of the boundary surface $\partial D$.
The effective action $\Gamma_E[D]$ in $d$ dimensions is given by,
\begin{equation}
\Gamma_E[{\bf x}]=\int_{\partial D}{\bf n}\cdot\boldsymbol{\alpha}\,dA-\lambda\int_D dV,
\label{Z2action}
\end{equation}
We also assume that under variations of the surface $\delta\boldsymbol{\alpha}\cdot{\bf n}=0$.

Using the divergence theorem we can recast the action in purely area form,
\begin{equation}
\Gamma_E[{\bf x}]=\int_{\partial D}{\bf n}\cdot\left(\boldsymbol{\alpha}-\frac{\lambda}{d}{\bf x}\right)dA,
\label{Z2action}
\end{equation}
When varying the action, we make use of the identity
\begin{equation}
\int_{\partial D}\delta{\bf n}\cdot{\bf x}\,dA=(d-1)\int_{\partial D}{\bf n}\cdot\delta{\bf x}\,dA
\end{equation}
The variation of the action becomes
\begin{equation}
\delta \Gamma_E=\int_{\partial D}\delta{\bf n}\cdot\left(\boldsymbol{\alpha}-\frac{\lambda}{d-1}{\bf x}\right)dA
\end{equation}
The instanton solution ${\bf x}_b$ and $\boldsymbol{\alpha}$ are therefore related by
\begin{equation}
\boldsymbol{\alpha}=\frac{\lambda}{d-1}{\bf x}_b+\gamma{\bf n}_b,
\end{equation}
where $\gamma$ is a constant (allowed since $\delta{\bf n}\cdot{\bf n}=0$). Substituting back into the
instanton action $\Gamma_E[{\bf x}_b]$ gives
\begin{equation}
\Gamma_E[{\bf x}_b]=\frac{\lambda}{d-1}\int_D dV+\gamma\int_{\partial D}dA
\end{equation}
We will take the simplest case $\gamma=0$.

It remains now to generalise the formulae for the surface tension vector quantity $\boldsymbol{\alpha}$.
We assume a generalisation of the previous expression in terms of hyperbolic functions,
\begin{equation}
\sum_{i=1}^d a_i\cosh\alpha_i=1
\end{equation}
The critical droplet now has a surface given by the equation
\begin{equation}
\sum_{i=1}^d a_i\cosh\left(\frac{\lambda x_i}{d-1}\right)=1\label{shape}
\end{equation}
The coefficients $a_i$ depend the couplings $J_i$, $i=1\dots d$. They should be symmetric under permutations
of the $J_i$. We also impose a consistent limit when $\delta\tau\to 0$, for $J_d=O(-\ln\delta\tau/2)$,
$J_i=O(\delta\tau)$ for $i\ne d$ and $\boldsymbol{\alpha}=O(\delta\tau)$. 
The simplest expression which has these properties, and reduces to the correct result for $d=2$, is
\begin{equation}
a_i=\frac{\tanh (2\sum_{j\ne i}J_j)}{(d-1)\cosh(2J_i)}.
\end{equation}

The quantum Ising model is obtained in the $\delta\tau\to0$ limit.
We make one final change, rescaling $\delta\tau$ in the expression for $J_d$,
\begin{equation}
J_d=\frac12\ln\coth\Gamma_c,\qquad \Gamma_c=\frac{gJ\delta\tau}{g_c}.
\end{equation}
The quantity $g_c$ will be identified as the value of $g$ at the critical point. 
The leading terms in $a_i$ are
\begin{equation}
a_{i\ne d}\approx\frac1{d-1}\left(1-2J_1^2-2\Gamma_c^2\right),
\quad a_d\approx 4\Gamma_c J_1.
\end{equation}
Inserting these terms into (\ref{shape}), with $\lambda=2hJ\delta\tau$ and $J_1=J\delta\tau$, 
predicts a circular bubble with equation for the radius $r(\tau)$,
\begin{equation}
r=\frac{(d-1)^{3/2}}{h}\left(1+\frac{g^2}{g_c^2}-2\frac{g}{g_c}\cosh\theta\right)^{1/2},\label{rad}
\end{equation}
and $\theta=2hJ\tau/(d-1)$. The action is particularly simple to evaluate in three dimensions,
\begin{equation}
\Gamma_E[{\bf x}_b]=2\int_0^{|\ln g/g_c|}\pi r^2d\theta=\frac{q}{h^2}
\end{equation}
where
\begin{equation}
q=16\pi\left[\left(1+\frac{g^2}{g_c^2}\right)\left|\ln\frac{g}{g_c}\right|
-\left|1-\frac{g^2}{g_c^2}\right|\right]
\end{equation}
Comparing this to the previous result (\ref{3Ddecay}) with $g\ll g_c$, we have an identical
factor of  $16\pi/h^2$ in the exponent.

Numerical simulations of false vacuum decay
have found that the rate has a similar form to the one quoted here, but with $\sim2\pi/h^2$ instead of
the $16\pi/h^2$ factor \cite{Pavesic:2026yiz}. This discrepancy is mysterious, but may indicate a rescaling of the $h$ coupling
in the opposite way to the rescaling that was applied to the $g$ coupling.

Notwithstanding some problems with these factors, we can construct an effective Hamiltonian for motion of the
bubble wall, introducing a parameter $\kappa=(d-1)^3$. Set
\begin{equation}
\omega(\theta)=\left(1+\frac{g^2}{g_c^2}-2\frac{g}{g_c}\cos\theta\right)^{1/2}
\end{equation}
Following the same steps as for the 1+1 dimensional case, we get the Hamiltonian
\begin{equation}
H=\kappa^{1/2} r\omega\left(\frac{p}{r}\right)-hr^2.
\end{equation}
The evolution of the bubble wall for an instanton nucleated event (\ref{rad}) is simply $r=\kappa^{1/2}\omega(hJt)/h$. However, the
non-critical bubble solutions with $H=E\ne0$ are rich and varied. We first solve for $\omega$,
\begin{equation}
\omega^2=\frac{(E+hr^2)^2}{\kappa r^2}.
\end{equation}
Hamilton's equations reduce to
\begin{equation}
\dot r^2+V=0,
\end{equation}
where
\begin{equation}
V=\frac{[\omega^2-(1+\hat g)^2][\omega^2-(1-\hat g)^2]\kappa^2}{4\omega^2},
\end{equation}
and $\hat g=g/g_c$. There are $0$, $2$ or $4$ roots of $V=0$ in the range $r>0$, and solutions oscillate
between pairs of roots.  

\section{Conclusion}

We have sought to explain how Coleman's ideas in false vacuum decay apply to quantum spin chains. The
treatment here has been elementary, largely pedagogical, and mostly based on results that date back several
decades.  However, the subject is developing rapidly, with new ideas in response to the prospects for laboratory tests of false
vacuum decay and overlaps with the theory of quantum annealing and quantum information.

The parameter ranges for which false vacuum decay with the spin chain resembles the field theory case is restricted, 
and is often complementary to regimes used for numerical simulations and current experiments. 
In particular, we have considered systems with a large total number of spins, and droplets containing 
many spins ($n_c\gg 1$). Since the decay rate is less than $\exp(-2n_c)$, Coleman's theory is mostly
applicable when the decay rates are very small. This would be doubly true for the two dimensional quantum spin chain,
where the conjectured decay rate has numerical factors in the exponent which are even larger than in the one 
dimensional case.

\bibliography{paper.bib}

\end{document}